\documentclass[twocolumn,showpacs,amsmath,amssymb,prb]{revtex4}

\usepackage{graphicx,rotating}% Include figure files
\usepackage{dcolumn}% Align table columns on decimal point
\usepackage{bm}% bold math
\usepackage{epsfig}
\usepackage{color}
\usepackage{soul}
\usepackage{array}
\usepackage{multirow}
\usepackage{subfigure}

\begin{document}

\title{$zT-$factor enhancement in SnSe: predictions from first principles calculations}

\author{Robert L. Gonz\'alez-Romero}\email{robertl2703@gmail.com}
\affiliation{Instituto de F\'isica Gleb Wataghin, Universidade Estadual de Campinas, S\~{a}o Paulo, 13083-859, Brazil}
\affiliation{Departamento de Sistemas F\'isicos, Qu\'imicos y Naturales. Universidad Pablo de Olavide. Ctra. de Utrera, km. 1, 41013, Sevilla, Spain}

\author{Alex Antonelli}\email{aantone@ifi.unicamp.br}
\affiliation{Instituto de F\'isica Gleb Wataghin, Universidade Estadual de Campinas, S\~{a}o Paulo, 13083-859, Brazil}

\author{Juan Jos\'e Mel\'endez}\email{melendez@unex.es}
\affiliation{Department of Physics, University of Extremadura, Avenida de Elvas, s/n, 06006, Badajoz, Spain}
\affiliation{Institute for Advanced Scientific Computing of Extremadura (ICCAEX), Avda. de Elvas, s/n. 06006, Badajoz, Spain}

\begin{abstract}
The electronic structure and thermoelectric properties of SnSe are studied by first-principles methods. The inclusion of van der Waals dispersive corrections improves the agreement of structural parameters with experiments. The bands structure and projected density of states justify the macroscopic anisotropy exhibited by this system. An original methodology is used to estimate the chemical potential and the relaxation time for the electrical and thermal conductivities. Following this methodology, the Seebeck coefficient and thermal conductivity for single crystals and polycrystals are described in good agreement with experimental data. As for the electrical conductivity, values calculated with a temperature-dependent relaxation time compare well with available measurements, especially for single crystals; polycrystals are better described by a constant relaxation time. Finally, the figure of merit of SnSe single crystals and polycrystals is calculated. It is found to exhibit a maximum for some ``ideal'' carrier concentration, and might be noticeably enhanced by using carrier concentrations higher than the experimental ones. From these findings, possible strategies to increase the figure of merit in practise are suggested.
\end{abstract}



\maketitle

\section{Introduction}

Thermoelectric materials have the capability to convert directly the residual heat resulting from industrial processes into electric energy. They are nowadays matter of great interest due to their applications in waste heat harvesting, radioisotope thermoelectric power generation, solid-state Peltier refrigeration, etc. \cite{Harman-02,Snyder-08,Qurashi-14,Poehler-16} From the point of view of potential applications, the main challenge for the scientific community consists on improving the thermoelectric properties of these materials to increase the efficiency of the heat conversion process. 

An ideal thermoelectric must have a high figure of merit $zT$, defined at a temperature $T$ as:
\begin{equation}
	\label{eq:fig_of_merit}
	zT = \frac {S^2\sigma T}{\kappa},
\end{equation}
where $S$, $\sigma$ and $\kappa$ are the Seebeck coefficient, the electrical and the thermal conductivities, respectively. The term $S^2\sigma$ is called power factor of the material. The thermal conductivity may be written as $\kappa=\kappa_L+\kappa_e$, where $\kappa_L$ denotes the lattice conductivity and $\kappa_e$ the contribution due to the charge carriers. The latter, on its own, is related to the electrical conductivity and the temperature through the Wiedemann-Franz law, $\kappa_e=T\sigma L$, where $L$ is a constant called Lorenz number. 
 
The search of materials with high $zT$ is an essential issue for possible technological applications. \cite{Vineis-10} According to Eq. (\ref{eq:fig_of_merit}), a high $zT$ requires a high power factor as well as a low thermal conductivity. Unfortunately, both conditions are difficult to achieve simultaneously because $S$ and $\sigma$ are strongly correlated in materials. Indeed, low carrier concentrations yield high Seebeck coefficients, but also low electric conductivities. On the other hand, a high electrical conductivity uses to be accompanied by a high thermal conductivity. Different optimization strategies have tried to reach an optimal equilibrium between these trends. \cite{Heremans-08, Biswas-12, Pei-11} Thus, since the 90’s, a myriad of thermoelectric materials with increasingly higher $zT$ factors, such as clathrates,\cite{Shi-10} skutterudites,\cite{Shi-11} germanium-silicon alloys, \cite{Joshi-08,Wang-08} systems with diamond-like structure\cite{Liu-12a} or the family of chalcogenides \cite{Heremans-08,Liu-12b,Tan-14,Zhao-14,Chen-14,Zhang-15,Sassi-14} have arisen as promising candidates for technological applications. Incidentally, we mention that an additional problem in this respect is the difficulty to reproduce the experimental research at the industrial scale. \cite{Heremans-08, Pei-11,Tan-14,Hinsche-12,Levi-14}.

The recent work by Zhao \textit{et al.} \cite{Zhao-14} about lead-free tin chalcogenides, more specifically those based on tin selenide (SnSe),\cite{Sassi-14,Heremans-14,Carrete-14,Yang-15,Wang-15,Guo-15,Ding-15,Gomes-15,Gharsallah-16,Zhang-16} has raised a great interest amongst the scientific community because SnSe is a simple compound that can be produced in a relatively easy way. The authors found an astonishingly low thermal conductivity in SnSe, giving rise to a very high and promising $zT=2.62 \pm 0.3$ along the $b$ axis at 923~K (typical values are $zT<2.0$ for skutterudites or $zT<1.5$ for clathrates \cite{Yang-16}). Independently, Chen \textit{et al.},\cite{Chen-14} Zhang \textit{et al.} \cite{Zhang-15} and Sassi \textit{et al.} \cite{Sassi-14} have reported high $zT$ values in SnSe polycrystals that, however, are lower than those reported by Zhao \textit{et al.} in single crystals. The main difference is the low thermal conductivity in single crystals, which has been ascribed to a strong anharmonicity in the bonds of SnSe.\cite{Zhao-14} On the other hand, Carrete \textit{et al.} \cite{Carrete-14} and Guo \textit{et al.}\cite{Guo-15} have carried out first-principles calculations to evaluate the lattice thermal conductivity of SnSe, and their results agree with experimental data by Chen \textit{et al.}\cite{Chen-14}, Zhang \textit{et al.}\cite{Zhang-15} and Sassi \textit{et al.}\cite{Sassi-14} A more recent work by Gharsallah \textit{et al.}\cite{Gharsallah-16} studies Ge-doped SnSe nanocrystals. Their experiments yield very high values of the Seebeck coefficient and the electrical resistivity, together with a very low thermal conductivity. The authors conclude that the grain boundaries decrease the thermal conductivity to values even below those reported for pure SnSe, but they are also likely to yield high electrical resistivities. 

In any case, these studies suggest that it is possible to develop thermoelectric materials with simple structures containing no lead but cheap and plentiful elements, in contrast with the currently available systems. To this end, understanding the physical origin of the discrepancies between the behavior of single crystals and polycrystals is a mandatory step towards the synthesis, and later commercialization, of more efficient thermoelectric materials based upon SnSe. This is the context of our work. We present an \textit{ab initio} study of the electronic structure and thermoelectric properties of SnSe. The calculated band structure sheds light on the thermal and transport coefficients of this system. As for the lattice thermal conductivity, we have used new computational tools that yield an excellent agreement with experimental data for $zT$. From this set of calculations, we have estimated optimal values of charge carrier concentrations that could maximize $zT$. We show then that the charge carrier concentration measured in experiments is far to be the ideal for an optimal $zT$ at moderate temperatures, and suggest possible ways to increase the figure of merit.
\section{Methodology}

\subsection{Computational setup}

\begin{figure}[h!]
	\includegraphics[width=0.8\columnwidth]{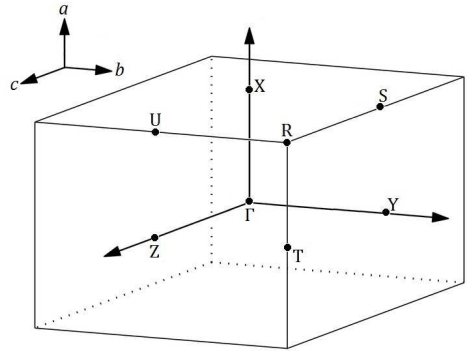}
	\caption{Schematic view of the first Brillouin zone for SnSe (not to scale). High-symmetry directions for bands calculations and crystal axes are shown. }
	\label{fig:BZ}
\end{figure}

First-principles calculations based on the density functional theory (DFT) \cite{Hohenberg-64,Kohn-65} have been performed using the projector augmented wave (PAW) method \cite{Blochl-94, Kresse-99} as implemented in the Vienna ab initio simulation package (VASP). \cite{Kresse-96} The Perdew-Burke-Ernzerhof functional for the generalized-gradient-approximation (GGA) was used to describe the electronic exchange–correlation interaction. \cite{Perdew-96} The kinetic energy cutoff of wave functions was set to 700 eV, and a Monkhorst-Pack\cite{Monkhorst-76}  $k-$mesh of $6 \times 18 \times 18$ was used to sample the first Brillouin zone (BZ) for integrations in the reciprocal space. A force less than 10$^{-6}\,$eV/\AA$\;$and a total change in energy less than 10$^{-8}\,$eV were selected as convergence criteria for the structural optimization. The optimization procedure explicitly included van der Waals interactions modeled by the DFT-D method by Grimme \textit{et al.} \cite{Grimme-10} The band structure and electronic density of states (DOS) were calculated at the optimized structure along high symmetry directions (namely $\Gamma -X$, $\Gamma-Y$, $\Gamma-Z$, $\Gamma-U$ and $\Gamma-S$) of BZ, shown in \textcolor{blue}{Figure \ref{fig:BZ}}.

The thermoelectric properties of SnSe were studied by solving the Boltzmann transport equation (BTE) within the relaxation time approximation (RTA). Under this approximation, the Cartesian components of the electrical conductivity, Seebeck coefficient and electronic thermal conductivity are written as
\begin{subequations}
	\begin{gather}
	\label{eq:sigma}
	\sigma_{ij}(\mu,T)=e^2\int \left(-\frac {\partial f_0(\varepsilon,T)}{\partial \varepsilon}\right) \Sigma_{ij}(\varepsilon) d\varepsilon \\
	\label{eq:seebeck}
	(\sigma S)_{ij}(\mu,T)=\frac eT \int \left(-\frac {\partial f_0(\varepsilon,T)}{\partial \varepsilon}\right)(\varepsilon- \mu) \Sigma_{ij}(\varepsilon) d\varepsilon \\
	\label{eq:kappa}
	(\kappa_e)_{ij}(\mu,T)= \frac 1T \int \left(-\frac {\partial f_0(\varepsilon,T)}{\partial \varepsilon}\right)(\varepsilon- \mu)^2 \Sigma_{ij}(\varepsilon) d\varepsilon 
\end{gather}
\end{subequations}

In Eqs. (\ref{eq:sigma}) to (\ref{eq:kappa}), $f_0(\varepsilon, T)$ holds for the Fermi-Dirac distribution function, $e$ and $T$ are, respectively, the electron charge and the temperature and $\Sigma_{ij}(\varepsilon)$ is the transport distribution function, defined as
\begin{equation}
	\Sigma_{ij}(\varepsilon)=\frac 1V \sum_{n,\vec k} v_i(n,\vec k)v_j(n,\vec k)\tau(n,\vec k)\delta(\varepsilon-\varepsilon_{n,\vec k}),
\end{equation}
where $V$ is the volume of the solid, $\varepsilon_{n,\vec k}$ is the energy of an electron in the $n-$th band at wave vector $\vec k$, $v_i (n,\vec k)$ is the $i-$th component of its velocity and the summation is over all bands and over the entire BZ. $\tau (n, \vec k)$ is the relaxation time for electrons, which depends on the electron state for each dispersing mechanism as well as on temperature. In what follows, we will accept that $\tau$ is a constant parameter within BZ, that is, independent on the electron wave vector. This has been shown to be a good approximation, even for anisotropic systems.\cite{Scheidemantel-03,Madsen-06} The chemical potential $\mu$ is a function of temperature as well, and it is related to the charge carrier concentrations in the solid. The transport coefficients were calculated with the BoltzWann post-processing code\cite{Pizzi-14} included in the Wannier90 package.\cite{Mostofi-08} For comparison with experiments, average values of the $\hat \sigma$, $\hat S$ and $\hat \kappa_e$ tensors were calculated as one third of the traces of the corresponding matrices. 

The lattice thermal conductivity $\kappa_L$ was calculated by solving BTE for phonons. The dispersion relations for phonons and the second-order harmonic interatomic force constants were calculated using the Phonopy package.\cite{Togo-08,Togo-15} In order to get reliable phonon spectra, a $2 \times 3 \times 3$ supercell was employed for force constants calculations. To obtain the third-order anharmonic force constants and to solve the BTE, the ShengBTE code was used. \cite{Li-14} A $2 \times 4 \times 4$ supercell was built to calculate the anharmonic forces, and the first-principles based real-space finite displacement difference approach was employed;\cite{Li-14} as for DFT, van der Waals corrections were used for these calculations. We chose a cutoff radius of 6.5 \AA, according to a previous work. \cite{Carrete-14} More details about this method can be found in Refs. \cite{Carrete-14,Li-14}

\subsection{Evaluation of the relaxation times}\label{method}

The transport coefficients are defined in terms of the electronic chemical potential $\mu$ and the relaxation time for the BTE, $\tau$. The main difficulty to estimate $\tau$ lies on the number and complexity of the scattering mechanisms active at a given temperature (phonons, impurities, grain boundaries and other structural defects, etc). Each of these is ruled by a particular temperature-dependent relaxation time; the overall value is then an average of the relaxation times for each active scattering mechanism. The usual procedure to estimate the relaxation time is to combine first-principles calculations with experimental results, as in Refs. \cite{Wang-15,Guo-15,Shi-15} This semi-empirical methodology evaluates the relaxation time at room temperature from comparison with experimental data, and subsequently uses it in calculations at different temperatures. 

As will be evident below, assessing thermoelectric behavior of SnSe requires a detailed evaluation of $\tau$ as a function of temperature for each type of material, namely single crystals or polycrystals. In this work, the relaxation times were estimated following a two-step procedure. First, we set the chemical potential which optimally fitted the Seebeck coefficient from a comparison between calculated and experimental data at each temperature. For this step one does not need any relaxation time, assumed to be constant within BZ, since the $\tau-$dependence is cancelled out by that of $\sigma$ in Eq. (\ref{eq:seebeck}). The second step consisted on comparing the experimental values for $\sigma$ with those calculated using the previous values of $\mu$. In this second step, $\tau$ is chosen so as to yield the best fit between experimental and calculated $\sigma$ at each temperature as well. Thus, after this second step one has the $\mu(T)$ and $\tau(T)$ functions yielding the best fits to experimental data. For single crystals and polycrystals, all the experimental data available to us were used to calculate the respective relaxation times.

\section{Results and discussion}

\subsection{Crystal structure}

\begin{figure}[h!]
	\includegraphics[width=\columnwidth]{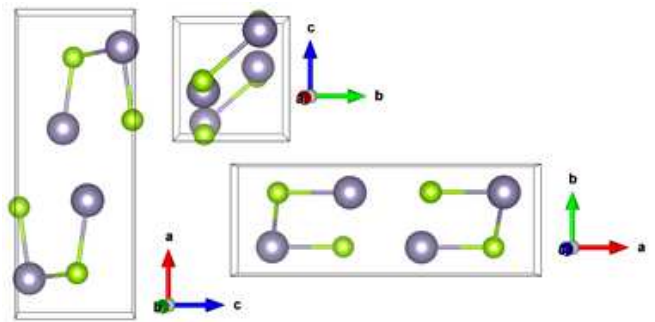}
	\caption{Crystal structure of SnSe in its orthorhombic room temperature phase along the three crystallographic directions.}
	\label{fig:cell}
\end{figure}

SnSe is orthorhombic at room temperature (space group $Pnma$, nr. 62), with the unit cell containing eight atoms arranged in two double adjacent layers, as shown in \textcolor{blue}{Figure \ref{fig:cell}}. The Sn atoms are surrounded by Se atoms in distorted octahedral coordination, forming a zig-zag arrangement.\cite{Zhao-14,Ding-15,Chattopadhyay-86,Li-15} As a preliminary step, we relaxed the unit cell of SnSe, since correct values of the lattice parameters are crucial to describe thermal and electrical properties. Previous studies show that DFT under the GGA overestimates the $a$ lattice parameter, and therefore also the separation between the layers, as shown in \textcolor{blue}{Table \ref{table:geometry}}. In addition, experimental evidence exists that the Sn-Se bonds are strong within the $BC$ planes but weak along the $a$ axis,\cite{Peters-90} suggesting that the latter could be due to van der Waals interactions. This is why we explicitly included van der Waals dispersive corrections during the cell relaxation stage. \textcolor{blue}{Table \ref{table:geometry}} shows the lattice parameters of the system after relaxation, showing an accurate agreement with experimental data, especially for the $a$ parameter. This is indicative that dispersive corrections are needed to get an appropriate description of the cell geometry in SnSe and of its physical behavior. The main disagreement is for the $c$ lattice parameter, for which DFT yields higher values than experiments. This result is not surprising, since the van der Waals interaction is expected to be negligible within the $BC$ plane; the difference between the calculated and the experimental values for $c$ is therefore that inherent to DFT under the GGA.

\begin{table}
	\renewcommand{\arraystretch}{1.4}
	\centering
	\caption{Lattice parameters (calculated with dispersive van der Waals corrections) and gap for SnSe. Calculated and experimental (marked with asterisks) values taken from the literature are included for comparison.}
	\vspace{0.3cm}
	\begin{tabular}{c cccc}
		\toprule 
		\textbf{Reference} & $a$ (\AA) & $b$ (\AA) & $c$ (\AA) & $\varepsilon_g$ (eV) \\
		\toprule
		This work & 11.56 & 4.17 & 4.54 & 0.63 \\ %\hline
		\multirow{2}{*}{Zhao \textit{et al.}\cite{Zhao-14}} & 11.79 & 4.21 & 4.55 & 0.61 \\ \cline{2-5}
		& 11.58$^*$ & 4.22$^*$ & 4.40$^*$ & 0.86$^*$ \\ %\hline
		Zhang \textit{et al.}\cite{Zhang-15} & 11.48 & 4.15 & 4.43 & 0.94 \\ %\hline
		Sassi \textit{et al.}$^*$\cite{Sassi-14} & 11.50 & 4.15 & 4.43 & --- \\ %\hline
		Carrete \textit{et al.}\cite{Carrete-14} & 11.72 & 4.20 & 4.55 & --- \\ %\hline
		Ding \textit{et al.}\cite{Ding-15} & 11.75 & 4.20 & 4.44 & 0.69 \\ %\hline		
		Gomes \textit{et al.}\cite{Gomes-15} & 11.81 & 4.22 & 4.47 & 1.00 \\ %\hline
		Gharsallah \textit{et al.}\cite{Gharsallah-16} & 11.54$^*$ & 4.16$^*$ & 4.45$^*$ & 0.58 \\ %\hline
		Chattopadhyay \textit{et al.}$^*$\cite{Chattopadhyay-86} & 11.50 & 4.15 & 4.45 & --- \\ %\hline
		Peng \textit{et al.}\cite{Peng-16} & 11.493$^*$ & 4.152$^*$ & 4.438$^*$ & $\approx 0.6$ \\ %\hline
		Singh \textit{et al.}$^*$\cite{Singh-91} & --- & --- & --- & 1.00 \\ %\hline
		Yu \textit{et al.}$^*$\cite{Yu-81} & --- & --- & --- & 0.923 \\ %\hline
		Albers \textit{et al.}$^*$\cite{Albers-62} & 11.51 & 4.13 & 4.5 & 0.9-0.95 \\ 
 \toprule
	\end{tabular}
	\label{table:geometry}
\end{table}

\subsection{Band structure and density of states}

\begin{figure}[h!]
	\centering
	\begin{subfigure}[]{\includegraphics[width=0.7\columnwidth]{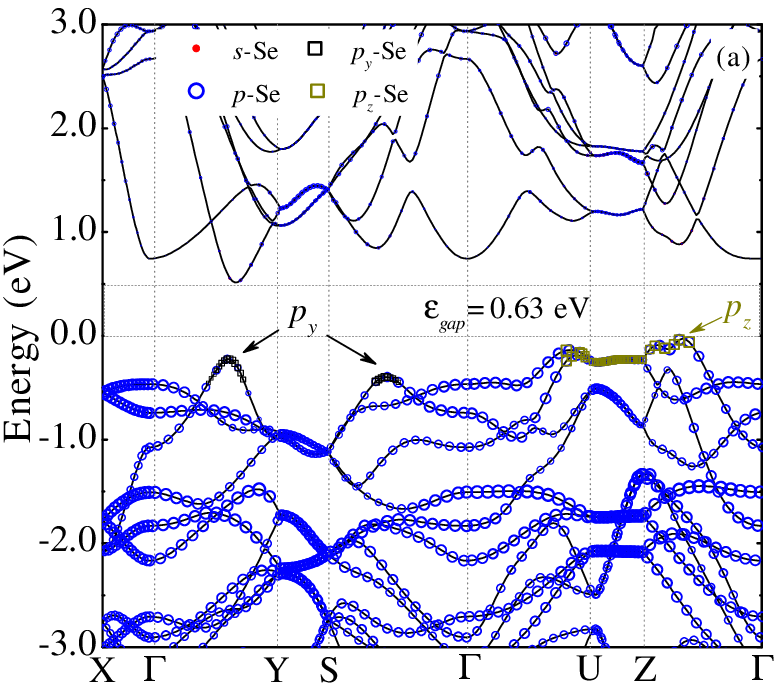}\label{subfig:bands_1}}\end{subfigure}
	\begin{subfigure}[]{\includegraphics[width=0.7\columnwidth]{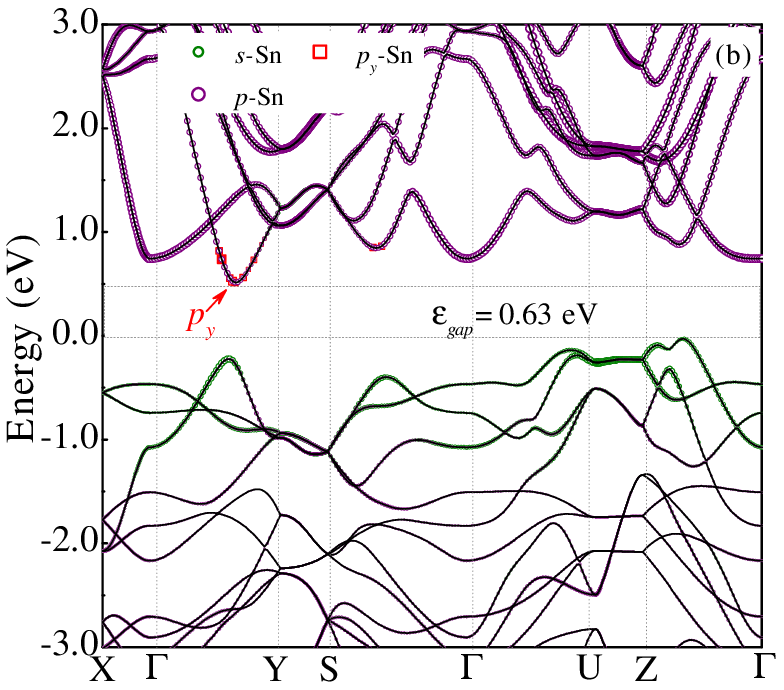}\label{subfig:bands_2}}\end{subfigure}
	\caption{Bands of SnSe projected onto $s$ and $p$ orbitals of Se (a) and Sn (b). The diameter of each circle is proportional to the partial density of states for a given energy. The Fermi energy was set to 0.0 eV. }
	\label{fig:bands}
\end{figure}

The physical properties of SnSe, which is a $p$-type semiconductor, are highly anisotropic.\cite{Zhao-14,Chen-14,Carrete-14,Yang-15,Guo-15,Ding-15,Shi-15} Such an anisotropy is particularly observed in measurements of $S$, $\sigma$ and $zT$, and may be understood in terms of the band structure of this material, shown in \textcolor{blue}{Figure \ref{fig:bands}}. Our calculations indicate that SnSe has an indirect gap of 0.63~eV (see \textcolor{blue}{Table~\ref{table:geometry}}), in good agreement with DFT values reported by other authors, but lower than the experimental gap (around 1.0 eV) due to the incapability of DFT to accurately describe the electronic exchange and correlation. The valence band maximum (VBM) lies on the $\Gamma-Z$ direction of BZ, parallel to the $c$ axis in the real space. There are several secondary maxima, with energies close to the main maximum, along the $\Gamma-Z$, $\Gamma-U$ and $\Gamma-Y$ directions, the later being parallel to the $b$ axis in the real space. The conduction band minimum (CBM) lies on the $\Gamma-Y$ direction, and there is a secondary minimum at $\Gamma$. The band structure of \textcolor{blue}{Fig.~\ref{fig:bands}} exhibits some ``pudding mold'' characteristics, as proposed by Kuroki and Arita.\cite{Kuroki-07} The systems with this type of bands use to have high electrical conductivities and Seebeck coefficients. 

\textcolor{blue}{Figure \ref{fig:bands}} plots the bands of SnSe projected onto $s$ and $p$ orbitals of Se (\textcolor{blue}{Fig. \ref{subfig:bands_1}}) and Sn (\textcolor{blue}{Fig. \ref{subfig:bands_2}}) as well. These projections appear as circles, whose diameters are proportional to the partial density of electronic states for the corresponding energy. The bands arisen from Se$-s$ states locate below -10.0 eV, and they do not appear in \textcolor{blue}{Fig. \ref{subfig:bands_1}}. The valence bands have mainly Se$-p$ character, with the VBM associated to the $p_z$ states with a small contribution of Sn$-s$ ones (shown in \textcolor{blue}{Fig. \ref{subfig:bands_2}}). The secondary maxima have Se$-p_z$ character too, except the one along the crystal axis $b$ ($\Gamma-Y$ direction), which arises from Se$-p_y$ states. On the contrary, the conduction bands are formed mostly by Sn$-p$ states. The CBM corresponds to Sn$-p_y$ states. Some valence bands have also Sn$-s$ character, although most of the bands arisen from Sn$-s$ levels have low energies (below -4.0 eV). It is also noticeable that both CBM and VBM lie on planes perpendicular to the $a$ axis, which justifies why the transport coefficients are so different along that direction.

\subsection{Seebeck coefficient and electrical conductivity}\label{seebeck}

The dependence of $S$ and $\sigma$ of SnSe with the charge carrier concentration and the temperature is still under study. \cite{Yang-15,Wang-15,Guo-15,Ding-15,Zhang-16,Shi-15,Guan-15} Most researchers agree that the experimental $S(T)$ curve exhibits a maximum beyond which it is decreasing; this behavior has been ascribed to bipolar conduction or to the excitation of positive and negative charge carriers.~\cite{Zhao-14,Wang-15,Shi-15} Within the RTA for BTE, both the Seebeck coefficient and the electrical conductivity depend on the chemical potential (i.e., on the carrier concentration). On the other hand, $S$ does not depend on the relaxation time, as long as this is constant throughout ZB, whereas $\sigma$ increases linearly with $\tau$. These theoretical trends are not always directly comparable with experiments.\cite{Yang-15,Wang-15,Guo-15,Ding-15,Zhang-16,Shi-15,Guan-15} As we will show below, the disagreement, especially in single-crystal samples, could be due to the use of a temperature-independent $\tau$. Instead, a temperature-dependent $\tau$ fits reasonably well the experimental results and allows predict the behavior of SnSe at different conditions. 

\begin{figure}[h!]
	\centering
	\includegraphics[width=0.8\columnwidth]{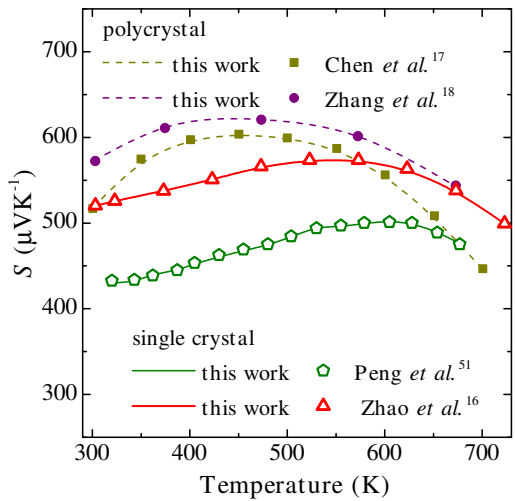}
	\caption{Calculated Seebeck coefficient of SnSe vs. temperature. Calculations were carried out for a temperature-dependent chemical potential.}
	\label{fig:seebeck}
\end{figure}

In this respect, \textcolor{blue}{Figure \ref{fig:seebeck}} shows the theoretical $S(T)$ curve calculated following the aforementioned methodology using a temperature-dependent chemical potential. To get these results, the entire conduction band was rigidly shifted upwards for the gap to equal the experimental value of 1.0 eV. The symbols in \textcolor{blue}{Fig. \ref{fig:seebeck}} correspond to experimental measurements carried out on two polycrystalline\cite{Chen-14,Zhang-15} and two single-crystal \cite{Zhao-14,Peng-16} samples. In all cases, $S$ increases monotonically with temperature to reach a maximum, beyond which it decreases. The temperature at maximum is around 600 K for single crystals, and below around 500 K for polycrystals. 

\textcolor{blue}{Fig. \ref{fig:seebeck}} exhibits a very good agreement between our calculations and experimental data for polycrystals as well as for single crystals. Besides, it shows that a careful choice of the chemical potential is required in order to calculate Seebeck coefficients in reasonable agreement with experimental data. In our case, the chemical potential was found to vary with temperature as the Seebeck coefficient does (see \textcolor{blue}{Supplement, Fig. S1}), although the temperature at maximum are slightly different. 

This inflection is caused by bipolar conduction effects. Indeed, the temperature-dependent chemical potential shown as \textcolor{blue}{Supplement, Fig. S1} yields an approximately constant electron concentration at temperatures up to around 550 - 600 K (see \textcolor{blue}{Supplement,~Figure S2}); at higher temperatures, the concentration of electrons begins to increase. Thus, one expects transport to be associated to holes below 550 K, the contribution from electrons becoming comparable above that temperature. At $T > 550$ K, the scattering mechanism become then more complex, and one should take into account the differences between the effective masses of electrons and holes, the interactions of both species with phonons and the possible electron-hole interactions. Similar bipolar effects have been reported elsewhere.\cite{Zhao-14,Wang-15,Shi-15}

\begin{figure}[h!]
	\centering
	\includegraphics[width=0.8\columnwidth]{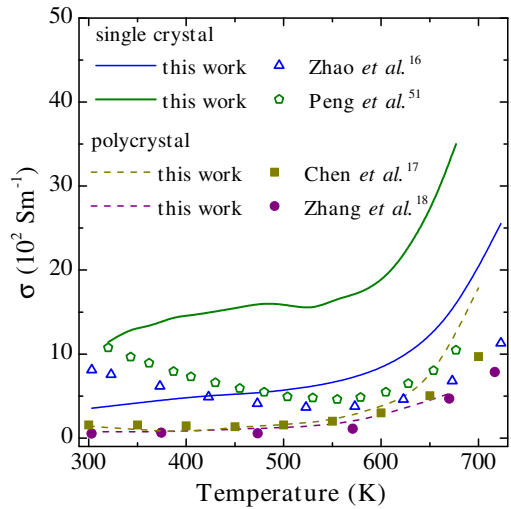}
	\caption{Calculated electrical conductivity of SnSe vs. temperature. At each temperature, calculations were performed using the same chemical potential as for the Seebeck coefficient (\textcolor{blue}{Fig. \ref{fig:seebeck}}).}
	\label{fig:conductivity}
\end{figure}

Once the $\mu(T)$ curve was properly calibrated, we tested our methodology as a predicting tool in SnSe. As an example, \textcolor{blue}{Figure~\ref{fig:conductivity}} plots the calculated electrical conductivity as a function of temperature. The conductivity data for polycrystals have been calculated for a constant relaxation time $\tau=4.0\cdot 10^{-15}$ s, which gives the best fit to the experimental data. As for the Seebeck coefficient, the agreement between experimental and calculated data is worst for single crystals when a constant relaxation time $\tau=10^{-14}$ s (which gives the best fit to experiments in this case) is used, however. Note that the relaxation time is higher for single crystals because at least one relaxation mechanism (by grain boundaries) must be absent from them.

\begin{figure}[h!]
	\centering
	\begin{subfigure}[]{\includegraphics[width=0.8\columnwidth]{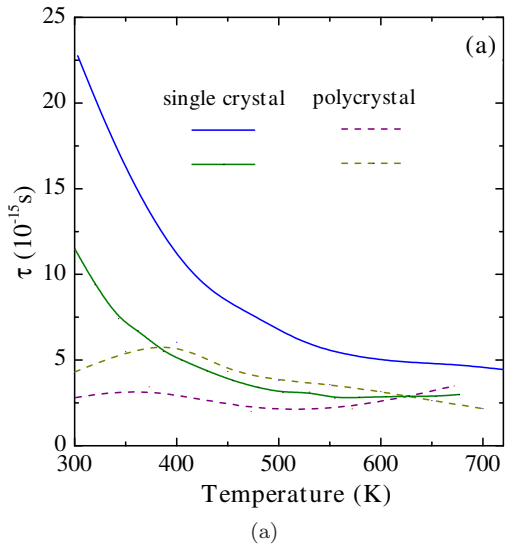}\label{subfig:tau}}\end{subfigure}
	\begin{subfigure}[]{\includegraphics[width=0.8\columnwidth]{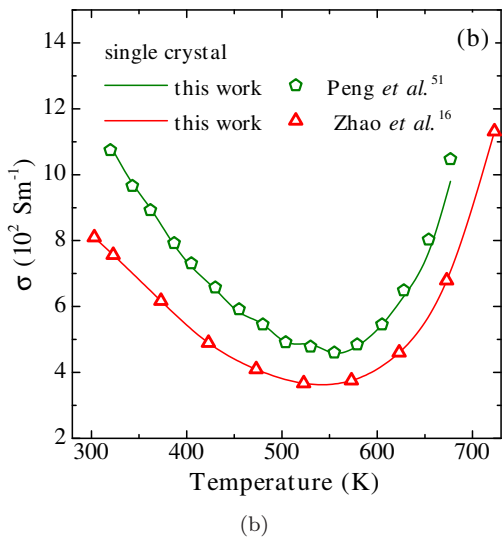}\label{subfig:conductivity}}\end{subfigure}
	\caption{a) Variation of the relaxation time with temperature for single crystals and polycrystals. Calculations were performed following the methodology explained in \textcolor{blue}{Sec.~\ref{method}}. b) Conductivity of single crystal SnSe calculated with the relaxation times of \textcolor{blue}{Fig. \ref{subfig:tau}}.}
\end{figure}

To improve the correspondence between calculated and experimental data, we moved to the second step of our methodology and considered a temperature-dependent relaxation time. The variation of $\tau$ with temperature is shown in \textcolor{blue}{Figure \ref{subfig:tau}}. For polycrystals, the calculated relaxation times decrease slowly with temperature. Thus, one may take an average relaxation time over the 300-700 K temperature range, and use this temperature-independent $\tau$ to compute the electrical conductivity in polycrystals. The good agreement of calculated data in \textcolor{blue}{Fig.~\ref{fig:conductivity}} with experiments validates this strategy, and shows that a constant relaxation time may be safely used for polycrystals. This approach differs from that used by other researchers. \cite{Wang-15,Ding-15,Guan-15} 

For single crystals, on the contrary, $\tau$ decreases significantly with temperature at low temperatures, and it reaches values comparable to those for polycrystals above about 550 K; a similar behavior has been reported by other authors. \cite{Biswas-11, Guo-15, Chen-14} This is not surprising, since conductivity at high temperature is mostly affected by scattering of the charge carriers by phonons, and this does not depend much on the structural details. The electrical conductivity of SnSe single crystals, computed from the relaxation times of \textcolor{blue}{Fig. \ref{subfig:tau}}, is shown in \textcolor{blue}{Figure \ref{subfig:conductivity}} as a function of temperature. In this case, contrarily to polycrystals, a temperature-dependent relaxation time yields a much better fit of the calculations to experimental data.

\subsection{Dependence on temperature of the relaxation time}

\begin{figure}[h!]
	\centering
	\includegraphics[width=0.8\columnwidth]{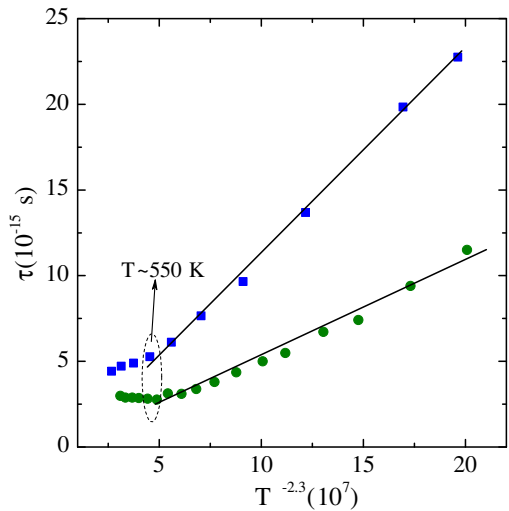}
	\caption{Plot of the relaxation time vs. $T^{-2.3}$, showing a linear dependence for $T > 550$ K.}
	\label{fig:fit_tau}
\end{figure}

The dependence of the relaxation time on temperature for single crystals deserves a more detailed study. Data from \textcolor{blue}{Fig. \ref{subfig:tau}} indicate that at temperatures between 300 K and 550 K, above which the bipolar effects become relevant, the relaxation time for single crystals is roughly proportional to $T^{-2.3}$, as shown in \textcolor{blue}{Figure \ref{fig:fit_tau}}. A temperature-dependent relaxation time is not surprising. Indeed, our DFT calculations were performed within the rigid-band approximation, where electron scattering, by either structural defects or phonons, is not considered. Consequently, all energy-dependent terms in Eqs. (\ref{eq:sigma}) to (\ref{eq:kappa}) are independent on temperature. Within the RTA, the possible electron scattering mechanisms are included as perturbations yielding a relaxation time depending on temperature. 

In a defect free system, electrons are expected to be scattered by acoustic phonons at low temperatures. The corresponding relaxation time is given by \cite{Jacoboni-10}
\begin{equation}
	\label{eq:tau_phonons}
	\tau(T)=\frac{2^{1/2}\pi \hbar^4\rho v_l^2}{3E^2(m^*k_BT)^{3/2}}\frac {F_0(\eta)}{F_{1/2}(\eta)},
\end{equation}

\noindent where $\rho$ is the density of the solid, $v_l$ is the longitudinal speed of sound, $E$, called strain potential, quantifies the strength of the electron-phonon interaction, $m^*$ is the effective mass of the electron and $\eta = \dfrac {\mu}{k_BT}$ is the reduced chemical potential. $F_{\alpha}(x)$ is the Fermi integral, defined as:
\begin{equation}
	\label{eq:fermi_integral}
	F_{\alpha}(x)=\frac 2{\sqrt \pi} \int_0^{\infty} \frac {y^{\alpha}}{e^{y-x}+1}dy
\end{equation}

Thus, one would expect a relaxation time varying with temperature as $\tau \propto T^{-1.5}$; on the contrary, the observed trend is $\tau \propto T^{-2.3}$, which has been widely reported in the literature. \cite{Biswas-11,Wei-14,Wang-12,Wang-14,Jaworski-09,Li-16,Ravich-13} A plausible explanation for this deviation could be to consider a temperature-dependent effective mass. \cite{Ravich-13} Such a $m^*(T)$ could be theoretically justified within a rigid-band approximation taking into account that Eq. (\ref{eq:tau_phonons}) is calculated from the cross section for electron-phonon scattering. In other words, Eq. (\ref{eq:tau_phonons}) arises from the fact that the features of the electron-phonon scattering, for whichever electronic state, vary with temperature. But the electronic states (i.e., the bands) themselves vary with temperature due to the electron-phonon interaction, thus yielding a temperature-dependent effective mass. The best fit to experimental data is achieved for $m^* \propto T^{0.5}$ within the 300 K - 550 K temperature range, and this results agrees well with data reported for other chalcogenides. \cite{Wang-12,Ravich-13,Androulakis-11,Zhou-14,Pei-12} Recently, Kutorasinski \textit{et al.} have argued about a possible temperature-dependent effective mass in SnSe, although the authors do not give any estimation for such a function. \cite{Kutorasinski-15}

At temperatures above 550 K one observes in \textcolor{blue}{Fig. \ref{fig:fit_tau}} a slope change which, in our opinion, is due to the bipolar effect mentioned in \textcolor{blue}{Sec. \ref{seebeck}}.

\subsection{Thermal conductivity}

The electronic thermal conductivity $\kappa_e$ varies with temperature likewise the electrical conductivity does. It remains roughly constant up to 550 K, approximately; at higher temperatures, it increases in average from roughly $5.0 \cdot 10^{-2}$ (at around 550 K) to $1.7\cdot10^{-1}$ Wm$^{-1}$K$^{-1}$ (at 700 K). These trends are the same for single crystals and for polycrystals (see \textcolor{blue}{Supplement,~Fig. S3}), but $\kappa_e$ is significantly smaller for the latter at low temperatures. 

\begin{figure}[h!]
	\centering
	\includegraphics[width=0.9\columnwidth]{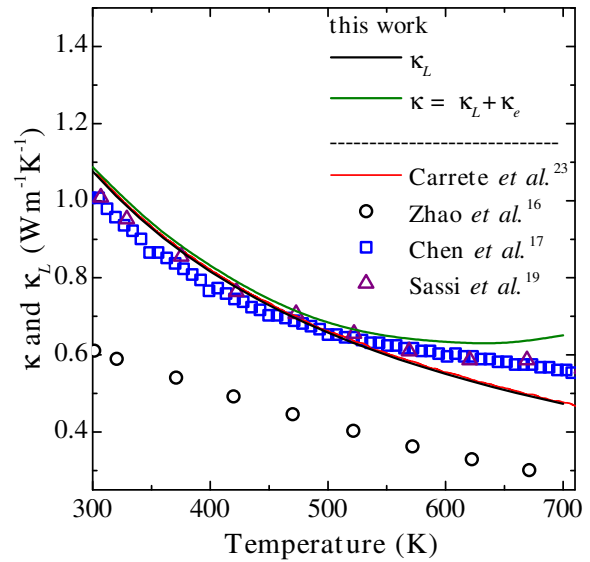}
	\caption{Temperature dependence of $\kappa_L$ and $\kappa$. Data taken from the literature are shown as reference.}
	\label{fig:kappa}
\end{figure}

At low temperature, the main contribution to the thermal conductivity is that from the lattice, $\kappa_L$. \textcolor{blue}{Figure \ref{fig:kappa}} plots our results for the lattice thermal conductivity as a function of temperature, averaged over the three crystal axes, together with some experimental and simulation results. \cite{Zhao-14,Chen-14,Sassi-14,Carrete-14} The agreement between our results and those those calculated by Carrete \textit{et al.},\cite{Carrete-14} is excellent, even though these authors do not include van der Waals corrections in their calculations. The thermal conductivity data by Chen \textit{et al.} \cite{Chen-14} and Sassi \textit{et al.} \cite{Sassi-14} are consistent with our results as well, especially at temperatures up to around 550 K; at higher temperatures, our $\kappa_L$ values are systematically lower. It is remarkable the low thermal conductivity measured by Zhao \textit{et al.},\cite{Zhao-14} especially considering that they used single-crystal samples. For these, one would expect the $\kappa_L$ values to be higher than in polycrystals, since the lack of grain boundaries should increase the thermal conductivity. \cite{Poudel-08,Dong-14}

\textcolor{blue}{Fig. \ref{fig:kappa}} plots the total (i.e., lattice plus electronic) thermal conductivity of SnSe as well. The difference between the total and lattice thermal conductivities is negligible at temperatures below 550 K, as one could expect. At higher temperatures, the electronic contribution to the total thermal conductivity becomes more and more important. At these temperatures, the inclusion of the electronic thermal conductivity improves the agreement between experimental and calculated data up to about 630~K.

\subsection{Figure of merit $zT$}

\begin{figure}[h!]
	\centering
	\includegraphics[width=0.9\columnwidth]{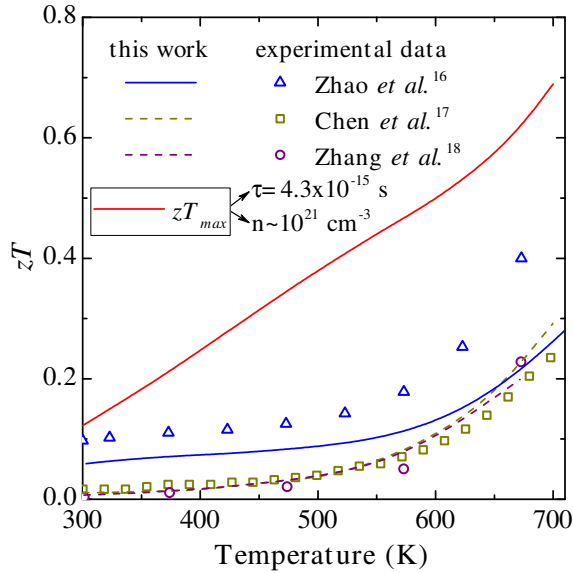}
	\caption{Figure of merit $zT$ as function of temperature. The red line corresponds to the $zT$ values calculated for an ``ideal'' carrier concentration.}
	\label{fig:zt}
\end{figure}

The previous results indicate that the different parameters involved in the calculations of transport coefficients, namely relaxation time and chemical potential, must be chosen very carefully if one wants to make predictions from first principles. This is particularly true for the figure of merit $zT$, since it describes the thermoelectric performance of materials. \textcolor{blue}{Figure \ref{fig:zt}} shows the figure of merit for SnSe calculated from Eq. (\ref{eq:fig_of_merit}) and the $S$ and $\sigma$ values reported in previous paragraphs. The reported $zT$ values for SnSe are usually low, ranging between less than 0.1 and 0.4 for $T <$ 700 K, approximately; this is the range of our results as well. For polycrystals, there is an excellent agreement between our results and those reported by Chen \textit{et al.} \cite{Chen-14} and Zhang \textit{et al.} \cite{Zhang-15} For single crystals, on the contrary, the agreement between our results and those by Zhao \textit{et al.} \cite{Zhao-14} is far to be satisfactory. In our opinion, this may be caused by the abnormally low lattice thermal conductivity found by the latter authors, which is about 50 \% lower than those calculated by us and other researchers (cf. \textcolor{blue}{Fig. \ref{fig:kappa}}).

\begin{figure}[h!]
	\centering
	\includegraphics[width=0.9\columnwidth]{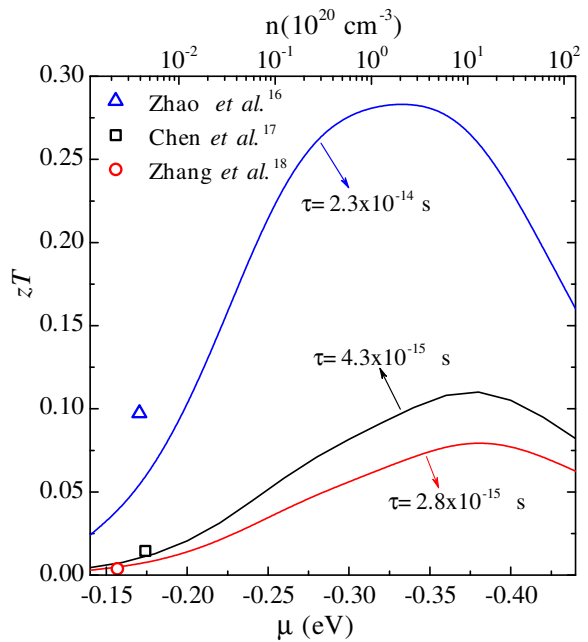}
	\caption{Figure of merit $zT$ as function of $\mu$ and the carrier concentration at 300 K. Experimental data are displayed as reference.}
	\label{fig:ideal_zt}
\end{figure}

In the context of thermoelectric performance, the question arises as to how to maximize the figure of merit $zT$. This is a difficult task in principle, since a correct balance between $S$ and $\sigma$, for which the carrier concentration plays a crucial role, is required. To evaluate the ``ideal'' carrier concentration (that is, the carrier concentration yielding the maximum $zT$) is complex as well. Fortunately, the versatility of the first principles calculations allows us to estimate the carrier concentration for which $zT$ is maximum. \textcolor{blue}{Figure \ref{fig:ideal_zt}} plots $zT$ as a function of the chemical potential (or, alternatively, the carrier concentration) at 300 K. To point out the strong effect of $\tau$ on $zT$ three different relaxation times, those estimated at 300 K for each sample used as reference (cf. \textcolor{blue}{Fig. \ref{subfig:tau}}), have been used to plot the data; for comparison, we include three experimental $zT$ values at 300~K. \cite{Zhao-14,Chen-14,Zhang-15} According to this plot, $zT$ reaches a maximum at the ``ideal'' carrier concentration, which depends strongly on the relaxation time. For instance, for $\tau=4.3 \cdot 10^{-15}$ s we get $zT_{max}=0.11$, and for $\tau=2.27 \cdot 10^{-14}$ s it is $zT_{max}=0.29$, approximately.

The optimal $zT$ values in \textcolor{blue}{Fig. \ref{fig:ideal_zt}} indicate that the carrier concentrations experimentally measured are much lower than the ``ideal'' ones predicted by calculations. Indeed, experimental carrier concentrations range between 10$^{17}$ and 10$^{18}$ cm$^{-3}$ at 300 K (see \textcolor{blue}{Supplement, Fig. S2}), whereas calculations predict it to be around 10$^{21}$ cm$^{-3}$ for polycrystals ($\tau=4.3 \cdot 10^{-15}$ s) and around 10$^{20}$~cm$^{-3}$ for single crystals ($\tau=2.27 \cdot 10^{-14}$ s). In other words, our results suggest that significantly higher $zT$ could be achieved at carrier concentrations higher than those commonly found in experiments, at least within the 350-650~K temperature range. An estimation of the maximum $zT$ appears in \textcolor{blue}{Fig. \ref{fig:zt}} (solid red line) as a function of temperature. For this plot, we chose a variable $\mu(T)$ set to yield $n=10^{21}$ cm$^{-3}$ at each temperature, and $\tau=4.3 \cdot 10^{-15}$ s, values estimated from data at 300~ K by Chen \textit{et al.} \cite{Chen-14} This finding may serve as a starting point for experimentalists seeking more efficient thermoelectric materials and opens two different ways for future research. One of them points to increase the relaxation times by handling the scattering mechanisms for charge carriers, whereas the second points to doping schemes able to increase the carrier concentrations towards values closer to the ``ideal'' ones.
\section{Conclusions}

The main conclusions of this work may be summarized as follows:

\begin{enumerate}
	\item The use of van der Waals dispersive corrections at DFT level leads to structural parameters that compare better with experimental data than those for uncorrected DFT. This is particularly important for the layer separation, which affects the physical behaviour.
	\item The band structure allows to understand the anisotropic behavior of SnSe, since marked differences are found in the bands curvatures along the cell axis $a$ with respect to axes $b$ and $c$.
	\item As for the thermoelectric properties, we show that the choice of the chemical potential and of the relaxation time affects critically the results. In polycrystals, the electrical conductivity may be properly described within the RTA with a constant relaxation time. On the contrary, an explicit temperature dependence of the relaxation time must be taken into account for single crystals. Incidentally, the grain boundaries are likely to greatly disperse the charge carriers, which would justify the little dependence of the relaxation time with the temperature in polycrystals.
	\item The thermal conductivity seems to be ruled by the lattice conductivity, at least at temperatures below about 550 K. At higher temperatures, the agreement between experiments and calculations improves if the carrier thermal conductivity is explicitly included. This change of trend at 550 K is related to bipolar conduction effects, which become relevant above that temperature. 
	\item A temperature-dependent relaxation time allows describe accurately the figure of merit of SnSe as well. For each relaxation time, it is found to exhibit a maximum that depends on the carrier concentration. In any case, carrier concentrations in the range 10$^{20}-10^{21}$ cm$^{-3}$ are likely to yield higher figures of merit than those currently achieved. This suggests that the relaxation time could be tailored by handling the proper dispersion mechanism for charge carriers, as well as using doping schemes yielding higher carrier concentrations
\end{enumerate}
\
% ========================================================
\begin{acknowledgments}
We gratefully acknowledge support from the Brazilian agencies CNPq, CAPES, the Center for Computational Engineering and Science-Fapesp/Cepid \#2013/08293-7 and FAPESP under Grant \#2013/14065-7. Support from the Junta of Extremadura, Spain, through Grant No. GR15105 is acknowledged as well. The calculations were performed at CCJDR-IFGWUNICAMP and at CENAPAD-SP in Brazil and at the University of Extremadura (Badajoz) in Spain.\\
\end{acknowledgments}

%\begin{appendix}
%\input{appA}
%\end{appendix}


\begin{thebibliography}{99}

\bibitem{Harman-02} T. C. Harman, P. J. Taylor, M. P. Walsh, B. E. LaForge, % Quantum dot superlattice thermoelectric materials and devices.
Science, {\bf 297}, 2229-2232 (2003).

\bibitem{Snyder-08} G. J. Snyder, E. S. Toberer, %Complex thermoelectric materials.
Nat. Mater., {\bf 7}, 105-114 (2008).

\bibitem{Qurashi-14} A. Qurashi, Metal chalcogenide nanostructures for renewable energy applications. John Wiley \& Sons, 2014. 

\bibitem{Poehler-16} T. O. Poehler, Innovative Thermoelectric Materials: Polymer, Nanostructure and Composite Thermoelectrics. World Scientific, 2016. 

\bibitem{Vineis-10} C. J. Vineis, A. Shakouri, A. Majumdar, M. G. Kanatzidis, %Nanostructured thermoelectrics:big efficiency gains from small features.
Adv. Mater., {\bf 22}, 3970-3980 (2010).

\bibitem{Heremans-08} J. P. Heremans, V. Jovovic, E. S. Toberer, A. Saramat, K. Kurosaki, A. Charoenphakdee, S. Yamanaka, G. J. Snyder, %Enhancement of thermoelectric efficiency in PbTe by distortion of the electronic density of states.
Science, {\bf 321}, 554-557 (2008).

\bibitem{Biswas-12} K. Biswas, J. He, I. D. Blum, C.-I. Wu, T. P. Hogan, D. N. Seidman, V. P. Dravid, M. G. Kanatzidis, %High-performance bulk thermoelectrics withh all-scale hierarchical architectures.
Nature, {\bf 489}, 414-418 (2012).

\bibitem{Pei-11} Y. Pei, X. Shi, A. LaLonde, H. Wang, L. Chen, G. J. Snyder, %Convergence of electronic bands for high performance bulk thermoelectrics.
Nature, {\bf 473}, 66-69 (2011).

\bibitem{Shi-10} X. Shi, J. Yang, S. Bai, J. Yang, H. Wang, M. Chi, J. R. Salvador, W. Zhang, L. Chen, W. Wong-Ng, %On the Design of High-Efficiency Thermoelectric Clathrates through a Systematic Cross-Substitution of Framework Elements.
Adv. Funct. Mater., {\bf 20}, 755-763 (2010).

\bibitem{Shi-11} X. Shi, J. Yang, J. R. Salvador, M. Chi, J. Y. Cho, H. Wang, S. Bai, J. Yang, W. Zhang, L. Chen, %Multiple-Filled Skutterudites: High Thermoelectric Figure of Merit through Separately Optimizing Electrical and Thermal Transports.
J. Am. Chem. Soc., {\bf 133}, 7837-7846 (2011).

\bibitem{Joshi-08} G. Joshi, H. Lee, Y. Lan, X. Wang, G. Zhu, D. Wang, R. W. Gould, D. C. Cuff, M. Y. Tang, M. S. Dresselhaus, G. Chen, Z. Ren, %Enhanced Thermoelectric Figure-of-Merit in Nanostructured $p-$type Silicon Germanium Bulk Alloys.
Nano Lett., {\bf 8}, 4670-4675 (2008). 

\bibitem{Wang-08} X. W. Wang, H. Lee, Y. C. Lan, G. H. Zhu, G. Joshi, D. Z. Wang, J. Yang, A. J. Muto, M. Y. Tang, J. Klatsky, S. Song, M. S. Dresselhaus, G. Chen, Z. F. Ren, %Enhanced thermoelectric figure of merit in nanostructured $n-$type silicon germanium bulk alloy.
App. Phys. Lett., {\bf 93}, 193121 (2008). 

\bibitem{Liu-12a} R. Liu, L. Xi, H. Liu, X. Shi, W. Zhang, L. Chen, %Ternary compound CuInTe$_2$: a promising thermoelectric material with diamond-like structure.
Chem. Commun., {\bf 48}, 3818-3820 (2012). 

\bibitem{Liu-12b} H. Liu, X. Shi, F. Xu, L. Zhang, W. Zhang, L. Chen, Q. Li, C. Uher, T. Day, G. J. Snyder, %Copper ion liquid-like thermoelectrics.
Nat. Mater., {\bf 11}, 422-425 (2012)

\bibitem{Tan-14} G. Tan, L.-D. Zhao, F. Shi, J. W. Doak, S.-H. Lo, H. Sun, C. Wolverton, V. P. Dravid, C. Uher, M. G. Kanatzidis, %High Thermoelectric Performance of $p-$Type SnTe via a Synergistic Band Engineering and Nanostructuring Approach.
J. Am. Chem. Soc., {\bf 136}, 7006-7017 (2014)

\bibitem{Zhao-14} L.-D. Zhao, V. P. Dravid, M. G. Kanatzidis, %Ultralow thermal conductivity and high thermoelectric figure of merit in SnSe crystals.
Nature, {\bf 508}, 373-377 (2014).

\bibitem{Chen-14} C.-L. Chen, H. Wang, Y.-Y. Chen, T. Day, G. J. Snyder, %Thermoelectric properties of $p-$type polycrystalline SnSe doped with Ag.
J. Mater. Chem. A, {\bf 2}, 11171-11176 (2014)

\bibitem{Zhang-15} Q. Zhang, E. K. Chere, J. Sun, F. Cao, K. Dahal, S. Chen, G. Chen, Z. Ren, %Studies on Thermoelectric Properties of $n‐$type Polycrystalline SnSe$_{1‐x}$S$_x$ by Iodine Doping.
Adv. Energy Mater., {\bf 5}, 1500360 (2015)

\bibitem{Sassi-14} S. Sassi, C. Candolfi, J.-B. Vaney, V. Ohorodniichuk, P. Masschelein, A. Dausher, B. Lenoir, %Assessment of the thermoelectric performance of polycrystalline $p-$type SnSe.
Appl. Phys. Lett., {\bf 104}, 212105 (2014)

\bibitem{Hinsche-12} N. F. Hinsche, B. Y. Yavorsky, M. Gradhand, M. Czerner, M. Winkler, J. K\"{o}nig, H. B\"{o}ttner, I. Martig, P. Zahn, %Thermoelectric transport in Bi$_2$Te$_3$/Sb$_2$Te$_3$ superlattices.
Phys. Rev. B, {\bf 86}, 085323 (2012)

\bibitem{Levi-14} B. G. Levi, %Simple compound manifests record-high thermoelectric performance.
Phys. Today, {\bf 67}, 14 (2014)

\bibitem{Heremans-14} J. P. Heremans, %Thermoelectricity: The ugly duckling. 
Nature, {\bf 508}, 327-328 (2014)

\bibitem{Carrete-14} J. Carrete, N. Mingo, S. Curtarolo, %Low thermal conductivity and triaxial phononic anisotropy of SnSe. 
Appl. Phys. Lett.,  {\bf 105}, 101907 (2014)

\bibitem{Yang-15} J. Yang, G. Zhang, G. Yang, C. Wang, Y. X. Wang, %Outstanding thermoelectric performances for both $p-$ and $n-$type SnSe from first-principles study. 
J. Alloys Compd., {\bf 644}, 615-620 (2015)

\bibitem{Wang-15} F. Q. Wang, S. Zhang, J. Yu, Q. Wang, %Thermoelectric properties of single-layered SnSe sheet. 
Nanoscale, {\bf 7}, 15962-15970 (2015)

\bibitem{Guo-15} R. Guo, X. Wang, Y. Kuang, B. Huang, %First-principles study of anisotropic thermoelectric transport properties of IV-VI semiconductor compounds SnSe and SnS. 
Phys. Rev. B, {\bf 92}, 115202, (2015)

\bibitem{Ding-15} G. Din, G. Gao, K. Yao, %High-efficient thermoelectric materials: The case of orthorhombic IV-VI compounds. 
Sci. Rep. {\bf 5}, 9567 (2015)

\bibitem{Gomes-15} L. C. Gomes, A. Carvalho, %Phosphorene analogues: Isoelectronic two-dimensional group-IV monochalcogenides with orthorhombic structure. 
Phys. Rev. B, {\bf 92}, 085406 (2015)

\bibitem{Gharsallah-16} M. Gharsallah, F. Serrano-S\'anchez, N. M. Nemes, F. J. Mompe\'an, M. T. Fern\'andez-D\'iaz, F. Elhalouani, J. A. Alonso, %Giant Seebeck effect in Ge-doped SnSe. 
Sci. Rep., {\bf 6}, 26774 (2016)

\bibitem{Zhang-16} L.-C. Zhang, G. Qin, W.-Z. Fang, H.-J. Cui, Q.-R. Zheng, Q.-B. Yan, G. Su, %Tinselenidene: a Two-dimensional Auxetic Material with Ultralow Lattice Thermal Conductivity and Ultrahigh Hole Mobility. 
Sci. Rep., {\bf 6}, 19830 (2016)

\bibitem{Yang-16} J. Yang, L. Xi, W. Qiu, L. Wu, X. Shi, L. Chen, J. Yang, W. Zhang, C. Uher, D. J. Singh, %On the tuning of electrical and thermal transport in thermoelectrics: an integrated theory–experiment perspective.
Npj Comput. Mater., {\bf 2}, 15015 (2016)

\bibitem{Hohenberg-64} P. Hohenberg, W. Kohn, %Inhomogeneous Electron Gas. 
Phys. Rev., {\bf 136}, B864-B871 (1964)

\bibitem{Kohn-65} W. Kohn, L. J. Sham, %Self-Consistent Equations Including Exchange and Correlation Effects. 
Phys. Rev., {\bf 140}, A1133-A1138 (1965)

\bibitem{Blochl-94} P. E. Bl\"{o}chl, %Projector augmented-wave method.
Phys. Rev. B: Condensed Matter, {\bf 50}, 17953-17979 (1994)

\bibitem{Kresse-99} G. Kresse, D. Joubert, %From ultrasoft pseudopotentials to the projector augmented-wave method. 
Phys. Rev. B: Condensed Matter, {\bf 59}, 1758-1775 (1999)

\bibitem{Kresse-96} G. Kresse, J. Furthmüller, %Efficient iterative schemes for ab initio total-energy calculations using a plane-wave basis set. 
Phys. Rev. B: Condensed Matter, {\bf 54}, 11169-11186 (1996)

\bibitem{Perdew-96} J. P. Perdew, K. Burke, M. Ernzerhof, %Generalized Gradient Approximation Made Simple. 
Phys. Rev. Lett., {\bf 77}, 3865-3868 (1996)

\bibitem{Monkhorst-76} H. J. Monkhorst, J. D. Pack, %Special points for Brillouin-zone integrations. 
Physical Review B: Condensed Matter, 1976. 13(12): p. 5188-5192.
Phys. Rev. B: Condensed Matter, {\bf 13}, 5188-5192 (1976)

\bibitem{Grimme-10} S. Grimme, J. Antony, S. Ehrlich, J. Krieg, %A consistent and accurate ab initio parametrization of density functional dispersion correction (DFT-D) for the 94 elements H-Pu. 
J. Chem. Phys., {\bf 132}, 154104 (2010)

\bibitem{Scheidemantel-03} T. J. Scheidemantel, C. Ambrosch-Draxl, T. Tronhauser, J. V Badding, J. O. Sofo, %Transport coefficients from first-principles calculations. 
Phys. Rev. B, {\bf 68}, 125210 (200)

\bibitem{Madsen-06} G. K. H. Madsen, %Automated Search for New Thermoelectric Materials: The Case of LiZnSb. 
J. Am. Chem. Soc., {\bf 128}, 12140-12146 (2006)

\bibitem{Pizzi-14} G. Pizzi, D. Volja, B. Kozinsky, M. Fornari, N. Marzari, %BoltzWann: A code for the evaluation of thermoelectric and electronic transport properties with a maximally-localized Wannier functions basis. 
Comput. Phys. Commun., {\bf 185} 422-429 (2014)

\bibitem{Mostofi-08} A. A. Mostofi, J. R. Yates, Y.-S. Lee, I. Souza, D. Vanderbilt, N. Marzari, %Wannier90: A tool for obtaining maximally-localised Wannier functions. 
Comput. Phys. Commun., {\bf 178}, 685-699 (2008)

\bibitem{Togo-08} A. Togo, F. Oba, I. Tanaka, %First-principles calculations of the ferroelastic transition between rutile-type and CaCl at high pressures. 
Phys. Rev. B: Condensed Matter, {\bf 78}, 134106 (2008)

\bibitem{Togo-15} A. Togo, I. Tanaka, %First principles phonon calculations in materials science. 
Scripta Mater., {\bf 108}, 1-5 (2015)

\bibitem{Li-14} W. Li, J. Carrete, N. A. Katcho, N. Mingo, %ShengBTE: A solver of the Boltzmann transport equation for phonons.
Comput. Phys. Commun., {\bf 185}, 1747-1758 (2014)

\bibitem{Shi-15} G. Shi, E. Kioupakis, %Quasiparticle band structures and thermoelectric transport properties of p-type SnSe. 
J. Appl. Phys., {\bf 117}, 065103 (2015)

\bibitem{Chattopadhyay-86} T. Chattopadhyay, J. Pannetier, H. G. von Schnering, %Neutron diffraction study of the structural phase transition in SnS and SnSe. 
J. Phys. Chem. Solids, {\bf 49}, 879-885 (1986)

\bibitem{Li-15} Y. Li, X. Shi, D. Ren, J. Chen, L. Chen, %Investigation of the Anisotropic Thermoelectric Properties of Oriented Polycrystalline SnSe. 
Energies, {\bf 8}, 6275-6285 (2015){}

\bibitem{Peters-90} M. J. Peters, L. E. McNeil, %High-pressure M\"ossbauer study of SnSe. 
Phys. Rev. B, {\bf 41}, 5893-5897 (1990)

\bibitem{Peng-16} K. Peng, X. Lu, H. Zhan, S. Hui, X. Tang, G. Wang, J. Dai, C. Uher, G. Wang, X. Zhou, %Broad temperature plateau for high ZTs in heavily doped p-type SnSe single crystals. 
Energy Environ. Sci., {\bf 9}, 454-460 (2016).

\bibitem{Singh-91} J. P. Singh, %Transport and optical properties of hot-wall-grown tin selenide films. 
J. Mater. Sci.: Materials in Electronics, {\bf 2}, 105-108 (1991)

\bibitem{Yu-81} J. G. Yu, A. S. Yue, O. M. Stafsudd Jr., %Growth and electronic properties of the SnSe semiconductor. 
J. Cryst. Growth, {\bf 52}, 248-252 (1981)

\bibitem{Albers-62} W. Albers, C. Haas, H. Ober, G. R. Schodder, J. D. Wasscher, %Preparation and properties of mixed crystals SnS$_(1−x)$Se$_x$. 
J. Phys. Chem. Solids, {\bf 23}, 215-220 (1962).

\bibitem{Kuroki-07} K. Kuroki, R. Arita, %``Pudding mold'' band drives large thermopower in Na x CoO2. 
J. Phys. Soc. Jpn., {\bf 76}, 083707 (2007)

\bibitem{Guan-15} X. Guan, P. Lu, L. Wu, L. Han, G. Liu, Y. Song, S. Wang, %Thermoelectric properties of SnSe compound.
J. Alloys Compd., {\bf 643}, 116-120 (2015).

\bibitem{Biswas-11} K. Biswas, J. He, Q. Zhang, G. Wang, C. Uher, V. P. Dravid, M. G. Kanatzidis, %Strained endotaxial nanostructures with high thermoelectric figure of merit. 
Nat. Chem., {\bf 3}, 160-166 (2011).

\bibitem{Jacoboni-10} C. Jacoboni, Theory of Electron Transport in Semiconductors: A Pathway from Elementary Physics to Nonequilibrium Green Functions. 
Springer Science \& Business Media, 2010.

\bibitem{Wei-14} T.-R. Wei, H. Wang, Z. M. Gibbs, C.-F. Wu, G. J. Snyder, J.-F. Li, %Thermoelectric properties of Sn-doped $p-$type Cu$_3$SbSe$_4$: a compound with large effective mass and small band gap.
J. Mater. Chem. A, {\bf 2}, 13527-13533 (2014).

\bibitem{Wang-12} H. Wang, Y. Pei, A. D. LaLonde, G. J. Snyder, %Weak electron–phonon coupling contributing to high thermoelectric performance in $n-$type PbSe.
P. Natl. Acad. Sci. USA, {\bf 109}, 9705-9709 (2012).

\bibitem{Wang-14} H. Wang, Z. M. Gibbs, Y. Takagiwa, G. J. Snyder, %Tuning bands of PbSe for better thermoelectric efficiency.
Energy Environ. Sci., {\bf 7}, 804-811 (2014).

\bibitem{Jaworski-09} C. M. Jaworski, V. Kulbachinskii, J. P. Heremans, %Resonant level formed by tin in Bi$_2$Te$_3$ and the enhancement of room-temperature thermoelectric power.
Phys. Rev. B, {\bf 80}, 233201 (2009).

\bibitem{Li-16} S.-L. Li, K. Tsukagoshi, E. Orgiu, P. Samori, %Charge transport and mobility engineering in two-dimensional transition metal chalcogenide semiconductors.
Chem. Soc. Rev., {\bf 45}, 118-151 (2016).

\bibitem{Ravich-13} I. I. Ravich, Semiconducting lead chalcogenides. Springer Science \& Business Media, 2013.

\bibitem{Androulakis-11} J. Androulakis, D.-Y. Chung, X. Su, L. Zhang, C. Uher, T. C. Hasapis, E. Hatzikraniotis, K. M. Paraskevopoulos, M. G. Kanatzidis, %High-temperature charge and thermal transport properties of the $n$-type thermoelectric material PbSe.
Phys. Rev. B, {\bf 84}, 155207 (2011).

\bibitem{Zhou-14} M. Zhou, Z. M. Gibbs, H. Wang, Y. Han, C. Xin, L. Li, G. J. Snyder, %Optimization of thermoelectric efficiency in SnTe: the case for the light band.
Phys. Chem. Chem. Phys., {\bf 16}, 20741-20748 (2014).

\bibitem{Pei-12} Y. Pei, A. D. LaLonde, H. Wang, G. J. Snyder, %Low effective mass leading to high thermoelectric performance.
Energy Environ. Sci., {\bf 5}, 7963-7969 (2012).

\bibitem{Kutorasinski-15} K. Kutorasinski, B. Wiendlocha, S. Kaprzyk, J. Tobola, %Electronic structure and thermoelectric properties of $n$- and $p$-type SnSe from first-principles calculations
Phys. Rev. B, {\bf 91}, 205201 (2015).

\bibitem{Poudel-08} B. Poudel, Q. Hao, Y. Ma, Y. Lan, A. Minnich, B. Yu, X. Yan, D. Wang, A. Muto, D. Vashaee, %High-thermoelectric performance of nanostructured bismuth antimony telluride bulk alloys.
Science, {\bf 320}, 634-638 (2008)

\bibitem{Dong-14} H. Dong, B. Wen, R. Melnik, %Relative importance of grain boundaries and size effects in thermal conductivity of nanocrystalline materials. 
Sci. Rep., {\bf 4}, 7037 (2014) 

\end{thebibliography}
\end{document}

\title{Supplementary material for ``$zT-$factor enhancement in SnSe: predictions from first principles calculations''}

\author{Robert L. Gonz\'alez-Romero}%\email{robertl2703@gmail.com}
\affiliation{Instituto de F\'isica Gleb Wataghin, Universidade Estadual de Campinas, S\~{a}o Paulo, 13083-859, Brazil}
\affiliation{Departamento de Sistemas Físicos, Químicos y Naturales, Universidad Pablo de Olavide. Ctra. de Utrera, km. 1, 41013, Sevilla, Spain}

\author{Alex Antonelli}%\email{aantone@ifi.unicamp.br}
\affiliation{Instituto de F\'isica Gleb Wataghin, Universidade Estadual de Campinas, S\~{a}o Paulo, 13083-859, Brazil}

\author{Juan Jos\'e Mel\'endez}%\email{melendez@unex.es}
\affiliation{Department of Physics, University of Extremadura, Avenida de Elvas, s/n, 06006, Badajoz, Spain}
\affiliation{Institute for Advanced Scientific Computing of Extremadura (ICCAEX), Avda. de Elvas, s/n. 06006, Badajoz, Spain}

\maketitle

\begin{figure}[h!]
	\centering
	\includegraphics[width=9cm]{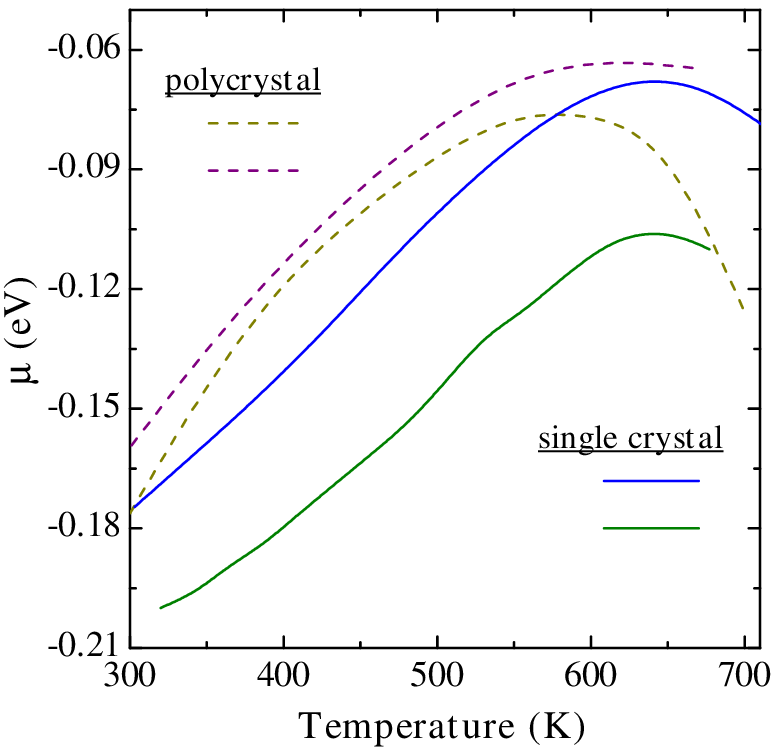} \\
	\bigskip
	Fig. S1: Variation of the chemical potential with temperature for SnSe single crystals and polycrystals.
	\label{fig:S1}
\end{figure}

This plot shows that the chemical potential for pure SnSe is negative, as one expects given its $p-$type character, and varies with temperature similarly for single crystals as for polycrystals. The maximum is reached at different temperatures, roughly 550 K and 600 K for polycrystals and single crystals, respectively.
 
\newpage

\begin{figure}[h!]
	\centering
	\includegraphics[width=9cm]{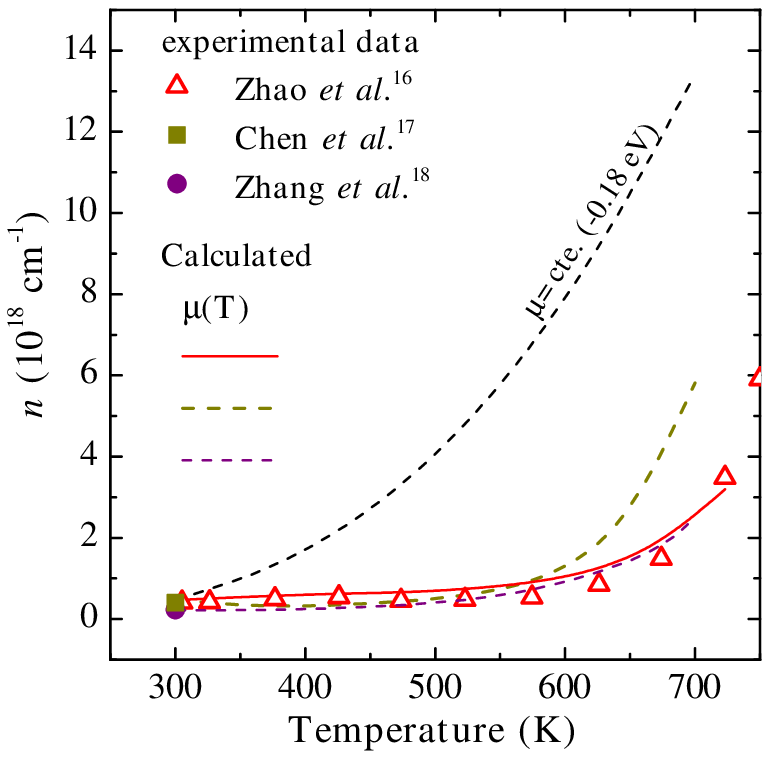} \\
	\bigskip
	Fig. S2: Concentration of electrons as function of temperature for several samples, calculated from the chemical potentials from Fig. S1.
	\label{fig:S2}
\end{figure}

This plot shows that the concentration of electrons in the conduction band remains practically constant in our calculations up to around 550 K, above which it starts to increase. This result, which is in very good agreement with experimental data, arises from the use of a temperature-dependent chemical potential. Indeed, a constant $\mu = $ 0.18eV yields a increasing concentration of electrons in the whole temperature range considered. In addition, its suggests that the change of trend observed in the transport coefficients may be due to bipolar effects. Reference numbers refer to the paper.
 
\newpage

\begin{figure}[h!]
	\centering
	\includegraphics[width=9cm]{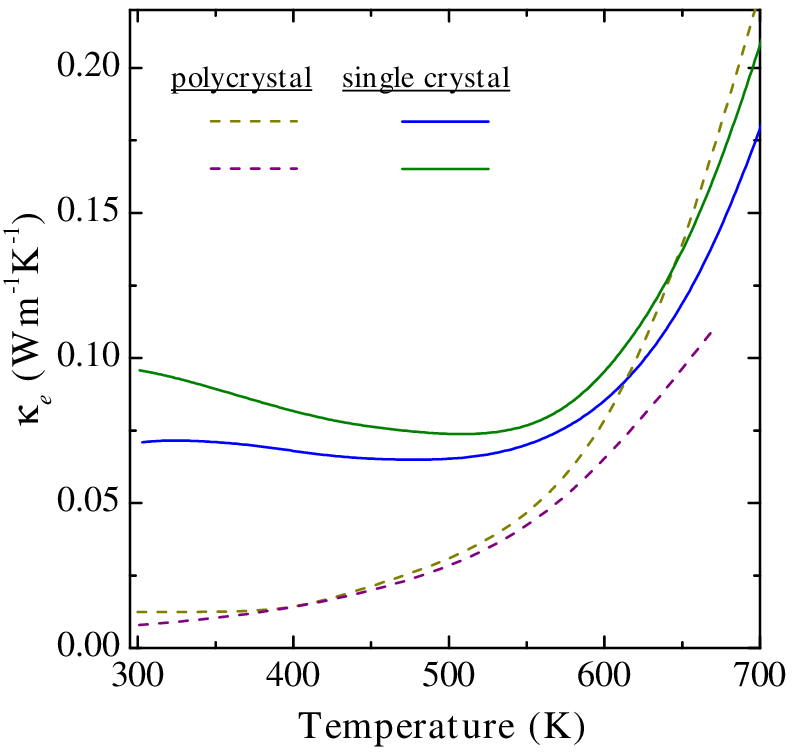} \\
	\bigskip
	Fig. S3: Electron thermal conductivities for SnSe single crystals and polycrystals.
	\label{fig:S3}
\end{figure}